\documentclass[conference]{IEEEtran}
\IEEEoverridecommandlockouts
\usepackage{cite}
\usepackage[left=1.72cm,right=1.72cm,top=0.71in,bottom=1.00in]{geometry}
\usepackage{amsmath,amssymb,amsfonts,mathtools,bbm}
\usepackage{physics}

\usepackage{algorithmic}
\usepackage{graphicx}
\usepackage{textcomp}
\usepackage{xcolor}
\usepackage{adjustbox}
\usepackage{graphicx}

\usepackage{tikz}
\tikzset{>=latex}

\usetikzlibrary{
    quantikz,
    shapes, 
    arrows,
    positioning
}

\usepackage{pgf}
\usepackage{pgfplots}
\pgfplotsset{compat=1.3}   
\usepgfplotslibrary{colorbrewer}

\PassOptionsToPackage{hyphens}{url}
\usepackage[bookmarks=false]{hyperref}
\usepackage{cite}
\usepackage[capitalize]{cleveref}

\usepackage{xcolor}
\usepackage[
    shortcuts,       
    acronym,         
    nomain,          
    section=chapter, 
    toc=true,        
]{glossaries}
\glsenableentrycount 

\definecolor{mypurple}{rgb}{0.49,0.18,0.56}
\definecolor{mygold}{rgb}{0.93,0.49,0.13}
\definecolor{mygreen}{rgb}{0,0.5,0}
\definecolor{myblue}{rgb}{0,0,0.75}
\definecolor{mymagenta}{cmyk}{0,1,0,0.12}
\definecolor{mygray}{rgb}{0.5,0.5,0.5}

\usepackage[T1]{fontenc}
\usepackage[utf8]{inputenc}

\usepackage{amsmath,amssymb,amsthm,mathtools}

\theoremstyle{remark}

\usepackage{hyperref,booktabs,graphicx,tikz}
\usepackage[capitalize]{cleveref}

\usepackage{amsfonts,bbm,braket}

\usepackage{adjustbox}

\usetikzlibrary{
quantikz, 
shapes,  
arrows,  
}

\DeclareUnicodeCharacter{0391}{\Alpha}       
\DeclareUnicodeCharacter{0392}{\gamma}        
\DeclareUnicodeCharacter{0393}{\Gamma}       
\DeclareUnicodeCharacter{0394}{\Delta}       
\DeclareUnicodeCharacter{0395}{\Epsilon}     
\DeclareUnicodeCharacter{0396}{\Zeta}        
\DeclareUnicodeCharacter{0397}{\Eta}         
\DeclareUnicodeCharacter{0398}{\Theta}       
\DeclareUnicodeCharacter{0399}{\Iota}        
\DeclareUnicodeCharacter{039A}{\Kappa}       
\DeclareUnicodeCharacter{039B}{\Lambda}      
\DeclareUnicodeCharacter{039C}{\Mu}          
\DeclareUnicodeCharacter{039D}{\Nu}          
\DeclareUnicodeCharacter{039E}{\Xi}          
\DeclareUnicodeCharacter{039F}{\Omicron}     
\DeclareUnicodeCharacter{03A0}{\Pi}          
\DeclareUnicodeCharacter{03A1}{\Rho}         
\DeclareUnicodeCharacter{03A3}{\Sigma}       
\DeclareUnicodeCharacter{03A4}{\Tau}         
\DeclareUnicodeCharacter{03A5}{\Upsilon}     
\DeclareUnicodeCharacter{03A6}{\Phi}         
\DeclareUnicodeCharacter{03A7}{\Chi}         
\DeclareUnicodeCharacter{03A8}{\Psi}         
\DeclareUnicodeCharacter{03A9}{\Omega}       
\DeclareUnicodeCharacter{03B1}{\alpha}       
\DeclareUnicodeCharacter{03B2}{\beta}        
\DeclareUnicodeCharacter{03B3}{\gamma}       
\DeclareUnicodeCharacter{03B4}{\delta}       
\DeclareUnicodeCharacter{03B5}{\epsilon}     
\DeclareUnicodeCharacter{03B6}{\zeta}        
\DeclareUnicodeCharacter{03B7}{\eta}         
\DeclareUnicodeCharacter{03B8}{\theta}       
\DeclareUnicodeCharacter{03B9}{\iota}        
\DeclareUnicodeCharacter{03BA}{\kappa}       
\DeclareUnicodeCharacter{03BB}{\lambda}      
\DeclareUnicodeCharacter{03BC}{\mu}          
\DeclareUnicodeCharacter{03BD}{\nu}          
\DeclareUnicodeCharacter{03BE}{\xi}          
\DeclareUnicodeCharacter{03BF}{\omicron}     
\DeclareUnicodeCharacter{03C0}{\pi}          
\DeclareUnicodeCharacter{03C1}{\rho}         
\DeclareUnicodeCharacter{03C3}{\sigma}       
\DeclareUnicodeCharacter{03C2}{\varsigma}    
\DeclareUnicodeCharacter{03C4}{\tau}         
\DeclareUnicodeCharacter{03C5}{\upsilon}     
\DeclareUnicodeCharacter{03D5}{\phi}         
\DeclareUnicodeCharacter{03C6}{\varphi}      
\DeclareUnicodeCharacter{03C7}{\chi}         
\DeclareUnicodeCharacter{03C8}{\psi}         
\DeclareUnicodeCharacter{03C9}{\omega}       

\usetikzlibrary{shapes.geometric, arrows}

\tikzstyle{startstop} = [rectangle, rounded corners, 
minimum width=3cm, 
minimum height=1cm,
text centered, 
draw=black, 
fill=red!30]

\tikzstyle{io} = [trapezium, 
trapezium stretches=true, 
trapezium left angle=70, 
trapezium right angle=110, 
minimum width=3cm, 
minimum height=1cm, text centered, 
draw=black, fill=blue!30]

\tikzstyle{process} = [rectangle, 
minimum width=3cm, 
minimum height=1cm, 
text centered, 
text width=3cm, 
draw=black, 
fill=orange!30]

\tikzstyle{decision} = [diamond, 
minimum width=3cm, 
minimum height=1cm, 
text centered, 
draw=black, 
fill=green!30]
\tikzstyle{arrow} = [thick,->,>=stealth]

\begin{document}

\title{A Review of the Applications of Quantum Machine Learning in Optical Communication Systems}

\author{%
    \IEEEauthorblockN{%
    Ark Modi\IEEEauthorrefmark{4}\IEEEauthorrefmark{1}(A.M.),
    Alonso Viladomat Jasso\IEEEauthorrefmark{4}\IEEEauthorrefmark{2}(A.V.J.),
    Roberto Ferrara\IEEEauthorrefmark{1}(R.F.),
    Christian Deppe\IEEEauthorrefmark{1}(C.D.),\\
    Janis Nötzel\IEEEauthorrefmark{2}(J.N),
    Fred Fung\IEEEauthorrefmark{3}(F.F.),
    Maximilian Schädler\IEEEauthorrefmark{3}(M.S.)
    }
    \IEEEauthorblockA{%
    \IEEEauthorrefmark{1}
    Institute for Communications Engineering (LNT),
    \\
    \IEEEauthorrefmark{2}
    Emmy Noether Group for Theoretical Quantum Systems Design
    \\
    Technical University of Munich, D-80333 Munich, Germany
    \\
    \IEEEauthorrefmark{3}
    Optical and Quantum Laboratory, Munich Research Center\\
    Huawei Technologies D\"usseldorf GmbH, Riesstr. 25-C3,80992 Munich, Germany
    \\
    \IEEEauthorrefmark{4}
    These authors contributed equally to the work
    \\
    Email:
    \{ark.modi, viladomat.jasso, roberto.ferrara, christian.deppe, janis.noetzel\}@tum.de
    \\
    \{fred.fung, maximilian.schaedler\}@huawei.com}
}

\maketitle

\begin{abstract}
    In the context of optical signal processing, quantum and quantum-inspired machine learning algorithms have massive potential for deployment. 
    One of the applications is in error correction protocols for the received noisy signals.
    In some scenarios, non-linear and unknown errors can lead to noise that bypasses linear error correction protocols that optical receivers generally implement.
    In those cases, machine learning techniques are used to recover the transmitted signal from the received signal through various estimation procedures.
    Since quantum machine learning algorithms promise advantage over classical algorithms, we expect that optical signal processing can benefit from these advantages.
    In this review, we survey several proposed quantum and quantum-inspired machine learning algorithms and their applicability with current technology to optical signal processing. 
\end{abstract}

\begin{IEEEkeywords}
Quantum Machine Learning, 
Quantum Algorithms,
Quantum Computing, 
6G Communication,
Quantum-Classical Hybrid Algorithms,
Optical Communication
\end{IEEEkeywords}

\section{Summary}

Artificial intelligence has made significant progress thanks to modern large-scale machine learning (henceforth referred to as ML), leading to the deployment of weakly intelligent cognitive systems in various aspects of daily and professional life. 
ML involves adjusting software agent parameters through training processes, allowing them to develop problem-solving skills. 
This progress relies on analyzing large amounts of task-specific training data to learn desired input-output behaviours. 
The success of modern ML is largely attributed to working with domain-agnostic models and training algorithms, with deep learning being especially successful. 
Deep learning utilizes artificial neural networks with billions of adjustable parameters, making them flexible and effective in various computational intelligence tasks. 
However, training deep neural networks requires vast amounts of representative data and considerable computational resources.
To train state-of-the-art systems effectively, like OpenAI's GPT-3, dedicated compute clusters and high-performance computing hardware are necessary due to the immense computational demands. 
As a result, the practical feasibility and success of current ML applications are highly dependent on access to such advanced computing resources.

Researchers are increasingly exploring quantum computing as a potential solution to the computational demands of modern ML systems. 
A ``quantum advantage" can manifest in various ways, primarily affecting time complexity or execution time, and accuracy. 
Quantum computing has made significant strides and promises faster computations in scientific and industrial applications. 
A number of works such as~\cite{Lloyd_HHL_2009, SchuldPetruccione2018, schuld2014, kerenidis2017recommendation, kerenidis2018q, lloyd2013quantum, Preskill_2018} claim to achieve a time advantage while works such as~\cite{IBM_QSVM_2019,IBM_QSVM_2020,stereo_paper_entropy,schuld-quantum-advantage,tiwari2020kernel} show accuracy and convergence gains.
Quantum computers operate on qubits, which exist in superposition and can carry more information than classical bits. 
Computation with qubits is probabilistic, and measurements cause decoherence, collapsing the qubit to a specific state. 
Quantum bits can be entangled, meaning the state of one qubit affects the state of others. 
We see (so far) that ``quantum'' advantage arises from these two key properties of quantum systems - entanglement and sampling.
Sampling advantage is noticeable in linear algebraic quantum machine learning (henceforth contracted to QML) procedures; however, in the Noisy Intermediate-Scale Quantum (NISQ) era, classical replication of this advantage is possible with only linear slowdowns~\cite{tangQ-PCA-2021}. 
The paradigm changes when Quantum Random Access Memory (QRAM) becomes available since it allows for the amortization of state preparation costs over multiple iterations, making QML more efficient.
On the other hand, entanglement, a quintessential quantum phenomenon, endows quantum systems with two primary advantages: 
(a) complex correlations: quantum entanglement enables the storage of intricate correlations within the data, facilitating the efficient representation of complex relationships in quantum machine learning models.
(b) quantum parallelism: entanglement allows for quantum parallelism, enabling the simultaneous processing of multiple data points or states, substantially accelerating some computations.
Entanglement allows quantum computers to work with exponentially larger search spaces, making them particularly useful for combinatorial optimization in artificial intelligence and certain ML techniques.

Adiabatic quantum computers, like those produced by D-Wave, are designed for solving combinatorial optimization problems known as QUBOs, which have applications in ML tasks like data clustering and support vector machine training.
Adiabatic quantum computing formulates problems as energy minimization tasks and uses Hamiltonian operators to find their ground states, which represent solutions to the problems. 
This approach utilizes the adiabatic theorem, allowing the system to transition from a known initial Hamiltonian to the problem Hamiltonian, finding the ground state in the process.
This method also benefits from quantum tunnelling, enabling it to overcome local minima and potentially solve problems exponentially faster than classical optimization. 
Adiabatic quantum computing shares similarities with the classical paradigm of Hopfield neural networks, making it a viable quantum analogue of this classical approach.

On the other hand, quantum gate computing manipulates qubits using quantum mechanical operators, resembling classical digital computing in terms of gates. 
Quantum circuits, composed of quantum gates, perform computations on qubits to achieve specific input/output behaviours.
Designing effective quantum circuits is a challenging task, and classical ML is increasingly used as a tool for quantum circuit design.
Quantum gate computing is attractive for ML due to its mathematical foundation in complex linear algebra and its ability to handle exponentially larger state spaces compared to classical computing. 
This makes it appealing for tasks involving large high-dimensional data vectors, as it is expected to offer quantum speedup and potentially solve intractable problems.

A QML ``mini-revolution'' has occurred recently, with numerous scientific reports proposing quantum circuits for various ML tasks, such as linear algebra routines, regression, and classification. 
Present-day approaches for QML involve variational quantum computing algorithms or hybrid quantum-classical methods. 
These methods use parameterized quantum circuits with tunable quantum gates and rely on classical optimization techniques to adjust the gate parameters for the desired computation.

Variational and hybrid quantum-classical algorithms are appealing because they reduce the quantum computing resources needed for successful QML.
Researchers also consider parameterized quantum circuits as a quantum analogue of classical deep neural networks, but there are important differences. 
Quantum gates implement unitary operators, not non-linear functions like in classical neural networks, and reading out the internal states of a quantum circuit destroys their quantum coherence. 
As a result, variational or hybrid quantum-classical algorithms are currently the primary approach for optimizing parameterized quantum circuits in QML.

The growing literature on QML shows promising potential for mainstream applications. 
However, it is essential to temper overly optimistic expectations, especially given the limitations of current NISQ computing.

Quantum algorithm design abstracts away the limitations of physical qubits present in current NISQ devices. 
These devices have less than a hundred qubits, limited coherence times, and low fault tolerance due to noise and fluctuations.
Creating and maintaining quantum states reliably over longer periods is challenging but expected to improve with technological advancements.
Promising candidates to look out for include superconducting qubits and topological qubits.  
Quantum error correction mechanisms, similar to those used in classical computing, are crucial for fault tolerance.

Present-day quantum computers face difficulties in handling large quantum circuits due to error-prone quantum gate operations. 
Adiabatic quantum computers offer more reliable manipulation of larger qubit systems, but they are limited to specific energy minimization problems and lack the universality of quantum gate computers.
Despite theoretical equivalence, emulating quantum circuits on adiabatic quantum computers requires unrealized qubit connectivity structures.

There are several practical limitations and challenges that need to be considered when applying quantum computing to ML tasks. They can be summarised as follows:

1. \emph{Encoding and decoding data}: Quantum algorithms require careful consideration of how classical data is encoded into quantum states and decoded back to classical representations. 
The effort for preparing quantum states and reading them back into classical memory greatly impacts the quantum advantage, especially if it becomes exponential.

2. \emph{Bit-level computing}: Present-day quantum computing is primarily focused on bit-level computing, lacking abstract data structures and control structures found in classical programming. 
This means that certain ML algorithms relying on these constructs are not realizable on current quantum computers.

3. \emph{Quantum compilers and APIs}: Efforts are being made to develop quantum compilers and high-level application programming interfaces (APIs) for quantum computing. 
However, these tools are still in their early stages, and users need to think at the linear algebraic level of quantum computing.

4. \emph{Simulated quantum processors}: Some APIs allow for efficient digital simulations of quantum information processing, but these simulations are based on universal quantum processors. 
Algorithms that work on simulated quantum computers may not necessarily work on existing physical quantum computers.

5. \emph{Probabilistic nature}: Quantum computations involving measurements are inherently probabilistic, requiring repeated runs to obtain results.
Any outcome needs to be interpreted in terms of expectations rather than deterministic outcomes.

Taking these limitations into account is crucial before making claims about the superiority of quantum algorithms in ML. 
Practical implementation challenges and the current state of quantum computing technology need careful consideration.

QML is a relatively new field, and its best practices and standards are still evolving. 
Similar to the early days of classical ML, QML is currently facing challenges related to verifiability and reproducibility.
In classical ML's early stages, practical results were often reported without disclosing implementation details, data collection or processing protocols, and experimental procedures, leading to issues with the validity of claimed capabilities.

Presently, the field of QML often exhibits similar omissions in reporting crucial details of practical results. 
There is a lack of transparency in the scientific literature, making it difficult to evaluate rigorously the methods and reproducibility of reported outcomes. 
While this might be somewhat acceptable for a nascent field, it is essential to consider that the performance of current QML methods may not scale or generalize well to larger or different application settings.

To establish trust and credibility in QML, it will be vital for researchers to adopt practices similar to those in classical ML, providing code, data, and experimental protocols in their publications. 
By doing so, the field can progress towards ensuring the reliability and reproducibility of reported results and accelerate its growth.

Despite the caveats and limitations of QML, recent technological progress justifies serious engagement with the topic.
While quantum computers and algorithms are not yet mature enough to impact ML practically, ongoing development and substantial investments suggest rapid improvement is likely.
As the underlying technology advances, viable solutions may emerge, leading to unexpected developments and disruptions.

However, potential risks related to QML have not received as much attention as their benefits.
Ethics, reliability, trustworthiness, and safety have been recognized as important topics in classical ML, but similar scrutiny is yet to be applied to QML. 
As QML may significantly impact artificial cognitive systems, it is crucial to assess potential security issues. 
Further studies investigating the reliability, vulnerability, and potential new forms of attacks or defense mechanisms for critical digital infrastructures in the context of QML are required.
The assessment of QML from a cybersecurity perspective and a determination of measures to address security challenges is required.

In summary, QML is a promising, cutting-edge, and complex field that requires further development and exploration to unlock its full potential. 
As with any transformative technology, it will take time, research, and advancements in hardware and algorithms to fully understand its capabilities and limitations.

\section{Application of QML in Optical Communication}

QML is an emerging field that combines quantum computing principles with ML algorithms to solve complex problems. 
When applied to optical communication systems, QML can offer several advantages and applications. 
Here are some of the key areas where QML can be beneficial in optical communication systems:

\begin{enumerate}
 
 \item  \emph{Channel Estimation and Equalization}: In optical communication, signal distortions can occur due to various factors like dispersion and noise. 
 QML techniques can be used to estimate and equalize the channel conditions, enabling more reliable data transmission and improved communication performance~\cite{Deville_2021}.

 \item  \emph{Fault Detection and Error Correction}: QML algorithms can be applied to detect and correct errors that arise during data transmission in optical communication systems. 
 This can enhance the overall reliability and robustness of the communication network~\cite{nonStereoPaper,stereo_paper_entropy}.

 \item  \emph{Optimal Resource Allocation}: QML can optimize the allocation of resources in optical communication networks, such as determining the best routes for quantum signals or optimizing the placement of quantum repeaters to extend the communication distance~\cite{Rekha2022, narottama2022quantum}.

 \item  \emph{Adaptive Photonics}: QML can be applied to adaptive photonics, where the properties of photons are optimized in real-time to maximize the communication performance. 
 This can lead to adaptive and self-optimizing optical communication systems~\cite{chabaud2021quantum}.
 
\end{enumerate}

As mentioned before, it is important to note that QML is still a developing area, and its practical implementation in optical communication systems is an active research field. 
Currently, most of the research output is theoretical and sometimes not in the NISQ context, making the determination of the state-of-the-art difficult. 
However a comparison with well-known methods for decoding M-QAM optical fibre signals is presented in~\cite{stereo_paper_entropy}.
As quantum technologies advance, the integration of QML with optical communication holds great promise for revolutionizing the way we transmit and process information.

\tikzstyle{block} = [rectangle, draw, text width=9em, text centered, rounded corners, minimum height=4em]
\begin{figure*}
	
    \centering
	\footnotesize
    
    \begin{center}
        \begin{adjustbox}{max width = \textwidth}
		  \begin{tikzpicture}[node distance=3cm, >=latex']

            \node (start) [block] 
            {Machine Learning applications in 
            optical communications};
            \node (l1-1) [block, below of=start] 
            {Proactive fault 
            detection};
            \node (l1-2) [block, left of=l1-1] 
            {Optical Performance
            Monitoring};
            \node (l1-3) [block, left of=l1-2] 
            {Nonlinearity
            compensation};
            \node (l1-4) [block, right of=l1-1] 
            {Software-defined 
            Networking};
            \node (l1-5) [block, right of=l1-4] 
            {Quality of 
            transmission
            estimation};
            \node (l1-6) [block, right of=l1-5] 
            {Physical layer
            design};
            \node (l2-1) [block, below of=l1-1]
            {
                $\bullet$ SVM with DES\\
                $\bullet$ SVM\\
                $\bullet$ Naive Bayes-based method\\
                $\bullet$ ANN \\
                $\bullet$ ANN with shape-based clustering
            };
            \node (l2-2) [block, below of=l1-2]
            {
            $\bullet$ ANN\\
            $\bullet$ Kernel-based methods\\
            $\bullet$ PCA\\
            $\bullet$ DNN\\
            $\bullet$ LSTM\\
            $\bullet$ CNN\\
            $\bullet$ SVM
            };
            \node (l2-3) [block, below of=l1-3]
            {
            $\bullet$ SVM \\ 
            $\bullet$ ELM\\
            $\bullet$ K-Means\\
            $\bullet$ EM algorithm\\
            $\bullet$ ANN\\
            $\bullet$ BP algorithm\\
            $\bullet$ DNN\\
            $\bullet$ DNN with PCA
            };
            \node (l2-4) [block, below of=l1-4]
            {
            $\bullet$ ANN\\
            $\bullet$ Q-learning\\
            $\bullet$ Q-network\\
            };
            \node (l2-5) [block, below of=l1-5]
            {
            $\bullet$ SVM\\
            $\bullet$ Logic Regression\\
            $\bullet$ kNN Clustering\\
            $\bullet$ ANN \\
            $\bullet$ Random forest\\
            };
            \node (l2-6) [block, below of=l1-6]
            {
            $\bullet$ DNN-based end-to-end system optimisation\\
            $\bullet$ ANN-based receiver for NFDM systems\\
            $\bullet$ ANN-based receiver for IM/DD systems\\
            $\bullet$ ANN-based Raman amplifier design\\
            };
            
            
            
            
            \draw [-latex] (start) -- (l1-1);
            \draw [-latex] (start) -- (l1-2);
            \draw [-latex] (start) -- (l1-3);
            \draw [-latex] (start) -- (l1-4);
            \draw [-latex] (start) -- (l1-5);
            \draw [-latex] (start) -- (l1-6);
            
            \draw [-latex] (l1-1) -- (l2-1);
            \draw [-latex] (l1-2) -- (l2-2);
            \draw [-latex] (l1-3) -- (l2-3);
            \draw [-latex] (l1-4) -- (l2-4);
            \draw [-latex] (l1-5) -- (l2-5);
            \draw [-latex] (l1-6) -- (l2-6);
            
        
        \end{tikzpicture}
      \end{adjustbox}
    \end{center}
    
\caption{ Source:~\cite{MLinOpticsBook}. Summary of Machine Learning applications in Optical Communication Systems.} 

\label{fig:MLapp}

\end{figure*}
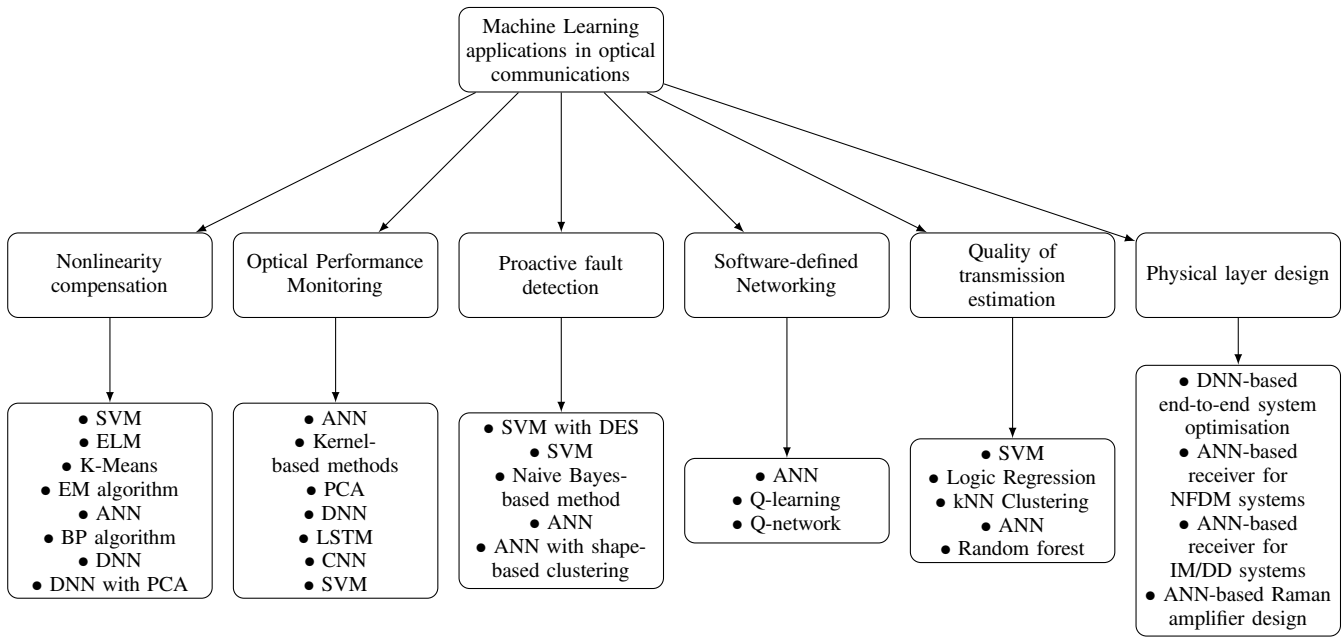

\cref{fig:MLapp} summarises the application of ML in optical communication systems. 
In general, in most applications where ML can be used, there exists a competing QML algorithm. 
As with any real-world application, the general algorithm has to be engineered, at an algorithmic, software and hardware level to achieve the best possible results.  
A case in point is k-means clustering - if one replaces the k-means algorithm with a hybrid quantum-classical implementation of the general quantum k-means algorithm proposed in~\cite{lloyd2013quantum}, the `advantage' is quite questionable, as shown in~\cite{nonStereoPaper}. 
However, as demonstrated in~\cite{stereo_paper_entropy}, with some engineering modifications and innovation, one can outperform the classical algorithm.
Another important thing to consider is the emerging class of quantum-inspired algorithms~\cite{germanSurvey, tangQ-PCA-2021, arrazola2020quantum, tang2019recommendation, frieze2004fast, drineas2006fast, achlioptas2007fast, mahoney2009cur, thurau2012deterministic, kannan2017randomized, salakhutdinov2007probabilistic, zhang2011quantum, kerenidis2017recommendation, chia2020quantum, gilyen2019quantum, chia2018quantum, Lloyd_HHL_2009}, which use methods and ideas inspired from quantum computing to optimise classical algorithms. 
For most of the applications mentioned in \cref{fig:MLapp}, there exist competing quantum or quantum-inspired methods that show promise in providing advantage over classical methods. Some of these algorithms are listed as follows: 
\begin{itemize}
    \item \emph{Neural Networks (NN) / Deep Learning} - Quantum Neural Networks (QNN)~\cite{hornik1989multilayer, rumelhart1986learning, lin2017does, coyleborn, liu2018differentiable, cheng2018information, allcock2018quantum, beer2020training, mitarai2018quantum, verdon2019learning, mcclean2018barren, zhu2019training, debnath2016demonstration, leyton2019robust, rossi2013quantum, zhao2019quantum, dayan1995helmholtz, hinton1995wake, vandam2020hybrid}
    
    \item \emph{Kernel-based methods} - Quantum kernels~\cite{tiwari2020kernel, schuld2019quantum, zhao2019bayesian, schuld2015simulating, IBM_QSVM_2019, bauckhage2017ising, schoelkopf2002learning, bauckhage2018adiabatic, schuld2014quantum, IBM_QSVM_2020} 
    
    \item \emph{Principal Component Analysis (PCA)} - Quantum PCA\cite{Lloyd_2014_PCA}, Quantum Inspired PCA~\cite{tangQ-PCA-2021,tang2019recommendation,arrazola2020quantum, chen2020quantum, xie2017quantum}
    
    \item \emph{Support Vector Machine (SVM)} - Quantum SVM (QSVM)~\cite{IBM_QSVM_2019,IBM_QSVM_2020, zahorodko2021comparisons, havenstein2018comparisons, rebentrost2014quantum, wang2014application, wittek2014supervised, boser1992training}
    
    \item \emph{Expectation-Maximisation Algorithm} - Quantum Expectation-Maximization Algorithm~\cite{QEM_Miyahara_2017, QEM_kerenidis2020quantum,QEM_Miyahara_2020,QEM_7798981}
    
    \item \emph{Classical kNN clustering} - Quantum and Quantum-inspired kNN clustering~\cite{stereo_paper_entropy,sergioli2018quantum,poggiali2022quantum,tangQ-PCA-2021}

    \item \emph{Logic regression} - Quantum Logic regression \cite{ning2023quantum, kim2019quantum, grant2018hierarchical, liu2018design}

    \item \emph{Random Forest} - Quantum and Quantum-inspired random forest algorithms \cite{khadiev2021quantum, srikumar2022kernel, xie2017quantum}
\end{itemize}

\section{Conclusions}
Quantum and Quantum-Inspired Machine Learning is an emerging field with significant potential and promise. 
However, like any nascent area of research, it faces several challenges and limitations. The following points describe the current state of QML.

\begin{itemize}

\item \emph{Early Stage of Development}: QML is still in its early stages, and researchers are actively exploring its potential applications and limitations. 
Many of the algorithms and techniques are still being developed and refined.

\item \emph{Hardware Limitations}: Building and maintaining quantum computers with a sufficient number of qubits and low decoherence rates is a very challenging task. 
As of now, quantum computers have not reached the level of efficiency and scalability to outperform classical computers for most ML tasks, especially when considering an industrial deployment.

\item \emph{Questionable Quantum Advantage}: The field of quantum computing often sees claims of "quantum advantage," which refers to the act of quantum computers solving problems faster than classical computers, or even solving problems that are practically infeasible for classical computers.
While there have been some demonstrations of quantum advantage in specific tasks such as prime factorisation, it is still a questionable claim in many cases, especially in the context of NISQ devices.
The source of this "quantum advantage" is also often unclear, and in some cases, the quantum ideas can be replicated classically to achieve great advantage - as can be seen in the case of sampling-based quantum-inspired algorithms. 

\item \emph{Interplay between Quantum and Classical}: Many ML tasks will always be performed efficiently using classical methods. 
QML is expected to have the most impact only in specific areas where quantum properties provide a computational advantage, such as solving certain optimization problems or simulating quantum systems.
Quantum systems can be used to compute and store complex correlations between data, and to parallelise certain computations -  this promises some advantage but the applications have to be engineered carefully. 

\item \emph{Algorithm Complexity}: Implementing and optimizing QML algorithms can be challenging and computationally expensive.

\item \emph{Data Requirements}: QML algorithms may require a large amount of high-quality quantum data, which is currently difficult to obtain.
Obtaining and preparing such data for QML tasks can be a significant hurdle.
The classical data loading problem, a result of the unavailability of stable quantum memory, is a significant issue that often introduces an exponential slowdown in the hybrid quantum-classical implementations of QML procedures. 

\end{itemize}

While QML faces these challenges, current research and technological development in this field are ongoing. 
As quantum computing technology, the potential for QML to impact various fields, including optimization, cryptography, and material science, remains a subject of active investigation.

\section*{Acknowledgement}

This work was funded by the TUM-Huawei Joint Lab on Algorithms for Short Transmission Reach Optics (ASTRO).
J.N. was funded from the DFG Emmy-Noether program under grant number NO 1129/2-1 and Munich Center for Quantum Science and Technology (MCQST).
C.D., and J.N. were funded by the Federal Ministry of Education and Research of Germany in the joint project 6G-life, project identification number: 16KISK002.
C.D., J.N., and A.V.J. were funded the Munich Quantum Valley (MQV) which is supported by the Bavarian state government with funds from the Hightech Agenda Bayern Plus.
C.D., J.N., and R.F. were funded by the Bavarian State Ministry for Economic Affairs, Regional Development and Energy in the project 6G and Quantum Technology (6GQT).
A.M., and C.D. were funded by the Federal Ministry of Education and Research of Germany in the project QR.X with the project number 16KISQ028.
We acknowledge useful discussions with Kareem H. El-Safty, and the use of ChatGPT 3.0 as a language tool. 
   
\bibliographystyle{IEEEtran}

\bibliography{biblio.bib}

\begin{thebibliography}{10}
\providecommand{\url}[1]{#1}
\csname url@samestyle\endcsname
\providecommand{\newblock}{\relax}
\providecommand{\bibinfo}[2]{#2}
\providecommand{\BIBentrySTDinterwordspacing}{\spaceskip=0pt\relax}
\providecommand{\BIBentryALTinterwordstretchfactor}{4}
\providecommand{\BIBentryALTinterwordspacing}{\spaceskip=\fontdimen2\font plus
\BIBentryALTinterwordstretchfactor\fontdimen3\font minus \fontdimen4\font\relax}
\providecommand{\BIBforeignlanguage}[2]{{%
\expandafter\ifx\csname l@#1\endcsname\relax
\typeout{** WARNING: IEEEtran.bst: No hyphenation pattern has been}%
\typeout{** loaded for the language `#1'. Using the pattern for}%
\typeout{** the default language instead.}%
\else
\language=\csname l@#1\endcsname
\fi
#2}}
\providecommand{\BIBdecl}{\relax}
\BIBdecl

\bibitem{Lloyd_HHL_2009}
\BIBentryALTinterwordspacing
A.~W. Harrow, A.~Hassidim, and S.~Lloyd, ``Quantum algorithm for linear systems of equations,'' \emph{Physical Review Letters}, vol. 103, no.~15, oct 2009. [Online]. Available: \url{https://doi.org/10.1103%2Fphysrevlett.103.150502}
\BIBentrySTDinterwordspacing

\bibitem{SchuldPetruccione2018}
\BIBentryALTinterwordspacing
M.~Schuld and F.~Petruccione, \emph{Supervised Learning with Quantum Computers}, ser. Quantum Science and Technology.\hskip 1em plus 0.5em minus 0.4em\relax Springer International Publishing, 2018. [Online]. Available: \url{https://books.google.de/books?id=1zpsDwAAQBAJ}
\BIBentrySTDinterwordspacing

\bibitem{schuld2014}
F.~P. M.~Schuld, I.~Sinayskiy, ``An introduction to quantum machine learning,'' \emph{arXiv:1409.3097 [quant-ph]}, 2014.

\bibitem{kerenidis2017recommendation}
\BIBentryALTinterwordspacing
I.~Kerenidis and A.~Prakash, ``{Quantum Recommendation Systems},'' in \emph{8th Innovations in Theoretical Computer Science Conference (ITCS 2017)}, ser. Leibniz International Proceedings in Informatics (LIPIcs), C.~H. Papadimitriou, Ed., vol.~67.\hskip 1em plus 0.5em minus 0.4em\relax Dagstuhl, Germany: Schloss Dagstuhl--Leibniz-Zentrum fuer Informatik, 2017, pp. 49:1--49:21. [Online]. Available: \url{http://drops.dagstuhl.de/opus/volltexte/2017/8154}
\BIBentrySTDinterwordspacing

\bibitem{kerenidis2018q}
I.~Kerenidis, J.~Landman, A.~Luongo, and A.~Prakash, ``q-means: A quantum algorithm for unsupervised machine learning,'' \emph{arXiv:1812.03584}, 2018.

\bibitem{lloyd2013quantum}
S.~Lloyd, M.~Mohseni, and P.~Rebentrost, ``Quantum algorithms for supervised and unsupervised machine learning,'' \emph{arXiv:1307.0411}, 2013.

\bibitem{Preskill_2018}
\BIBentryALTinterwordspacing
J.~Preskill, ``Quantum computing in the {NISQ} era and beyond,'' \emph{Quantum}, vol.~2, p.~79, aug 2018. [Online]. Available: \url{https://doi.org/10.22331%2Fq-2018-08-06-79}
\BIBentrySTDinterwordspacing

\bibitem{IBM_QSVM_2019}
\BIBentryALTinterwordspacing
V.~Havl{\'{\i}}{\v{c}}ek, A.~D. C{\'{o}}rcoles, K.~Temme, A.~W. Harrow, A.~Kandala, J.~M. Chow, and J.~M. Gambetta, ``Supervised learning with quantum-enhanced feature spaces,'' \emph{Nature}, vol. 567, no. 7747, pp. 209--212, mar 2019. [Online]. Available: \url{https://doi.org/10.1038%2Fs41586-019-0980-2}
\BIBentrySTDinterwordspacing

\bibitem{IBM_QSVM_2020}
\BIBentryALTinterwordspacing
J.-E. Park, B.~Quanz, S.~Wood, H.~Higgins, and R.~Harishankar, ``Practical application improvement to quantum svm: theory to practice,'' 2020. [Online]. Available: \url{https://arxiv.org/abs/2012.07725}
\BIBentrySTDinterwordspacing

\bibitem{stereo_paper_entropy}
\BIBentryALTinterwordspacing
A.~Viladomat~Jasso, A.~Modi, R.~Ferrara, C.~Deppe, J.~Nötzel, F.~Fung, and M.~Schädler, ``Quantum and quantum-inspired stereographic k nearest-neighbour clustering,'' \emph{Entropy}, vol.~25, no.~9, 2023. [Online]. Available: \url{https://www.mdpi.com/1099-4300/25/9/1361}
\BIBentrySTDinterwordspacing

\bibitem{schuld-quantum-advantage}
\BIBentryALTinterwordspacing
M.~Schuld and N.~Killoran, ``Is quantum advantage the right goal for quantum machine learning?'' \emph{PRX Quantum}, vol.~3, p. 030101, Jul 2022. [Online]. Available: \url{https://link.aps.org/doi/10.1103/PRXQuantum.3.030101}
\BIBentrySTDinterwordspacing

\bibitem{tiwari2020kernel}
P.~Tiwari, S.~Dehdashti, A.~K. Obeid, M.~Melucci, and P.~Bruza, ``Kernel method based on non-linear coherent state,'' 2020.

\bibitem{tangQ-PCA-2021}
\BIBentryALTinterwordspacing
E.~Tang, ``Quantum principal component analysis only achieves an exponential speedup because of its state preparation assumptions,'' \emph{Phys. Rev. Lett.}, vol. 127, p. 060503, 2021. [Online]. Available: \url{https://link.aps.org/doi/10.1103/PhysRevLett.127.060503}
\BIBentrySTDinterwordspacing

\bibitem{Deville_2021}
\BIBentryALTinterwordspacing
Y.~Deville and A.~Deville, ``New single-preparation methods for unsupervised quantum machine learning problems,'' \emph{{IEEE} Transactions on Quantum Engineering}, vol.~2, pp. 1--24, 2021. [Online]. Available: \url{https://doi.org/10.1109%2Ftqe.2021.3121797}
\BIBentrySTDinterwordspacing

\bibitem{nonStereoPaper}
A.~Modi, A.~V. Jasso, R.~Ferrara, C.~Deppe, J.~Noetzel, F.~Fung, and M.~Schaedler, ``Testing of hybrid quantum-classical k-means for nonlinear noise mitigation,'' \emph{arXiv preprint arXiv:2308.03540}, 2023.

\bibitem{Rekha2022}
\BIBentryALTinterwordspacing
S.~Rekha, D.~Saxena, A.~K. Singh, and C.~N. Lee, ``{A Quantum Machine Learning driven Reliable Resource Allocation Model for Sustainable Cloud Data Center},'' 11 2022. [Online]. Available: \url{https://www.techrxiv.org/articles/preprint/A_Quantum_Machine_Learning_driven_Reliable_Resource_Allocation_Model_for_Sustainable_Cloud_Data_Center/21571932}
\BIBentrySTDinterwordspacing

\bibitem{narottama2022quantum}
B.~Narottama and T.~Q. Duong, ``Quantum neural networks for optimal resource allocation in cell-free mimo systems,'' in \emph{GLOBECOM 2022-2022 IEEE Global Communications Conference}.\hskip 1em plus 0.5em minus 0.4em\relax IEEE, 2022, pp. 2444--2449.

\bibitem{chabaud2021quantum}
U.~Chabaud, D.~Markham, and A.~Sohbi, ``Quantum machine learning with adaptive linear optics,'' \emph{Quantum}, vol.~5, p. 496, 2021.

\bibitem{MLinOpticsBook}
\BIBentryALTinterwordspacing
F.~N. Khan, C.~Lu, and A.~P.~T. Lau, ``Machine learning methods for optical communication systems,'' in \emph{Advanced Photonics 2017 (IPR, NOMA, Sensors, Networks, SPPCom, PS)}.\hskip 1em plus 0.5em minus 0.4em\relax Optica Publishing Group, 2017, p. SpW2F.3. [Online]. Available: \url{https://opg.optica.org/abstract.cfm?URI=SPPCom-2017-SpW2F.3}
\BIBentrySTDinterwordspacing

\bibitem{germanSurvey}
\BIBentryALTinterwordspacing
C.~Bauckhage \emph{et~al.}, ``Quantum machine learning state of the art and future directions.'' [Online]. Available: \url{https://www.bsi.bund.de/SharedDocs/Downloads/DE/BSI/Publikationen/Studien/QML/Quantum_Machine_Learning.pdf?__blob=publicationFile&v=10}
\BIBentrySTDinterwordspacing

\bibitem{arrazola2020quantum}
\BIBentryALTinterwordspacing
J.~M. Arrazola, A.~Delgado, B.~R. Bardhan, and S.~Lloyd, ``Quantum-inspired algorithms in practice,'' \emph{{Quantum}}, vol.~4, p. 307, 2020. [Online]. Available: \url{https://doi.org/10.22331/q-2020-08-13-307}
\BIBentrySTDinterwordspacing

\bibitem{tang2019recommendation}
\BIBentryALTinterwordspacing
E.~Tang, ``A quantum-inspired classical algorithm for recommendation systems,'' in \emph{Proceedings of the 51st Annual ACM SIGACT Symposium on Theory of Computing}, ser. STOC 2019.\hskip 1em plus 0.5em minus 0.4em\relax New York, NY, USA: Association for Computing Machinery, 2019, p. 217–228. [Online]. Available: \url{https://doi.org/10.1145/3313276.3316310}
\BIBentrySTDinterwordspacing

\bibitem{frieze2004fast}
\BIBentryALTinterwordspacing
A.~Frieze, R.~Kannan, and S.~Vempala, ``Fast monte-carlo algorithms for finding low-rank approximations,'' \emph{Journal of the ACM}, vol.~51, no.~6, 2004. [Online]. Available: \url{https://www.math.cmu.edu/~af1p/Texfiles/SVD.pdf}
\BIBentrySTDinterwordspacing

\bibitem{drineas2006fast}
\BIBentryALTinterwordspacing
P.~Drineas, R.~Kannan, and M.~Mahoney, ``Fast monte carlo algorithms iii: Computing a compressed approximate matrix decomposition,'' \emph{SIAM Journal on Computing}, vol.~36, no.~1, 2006. [Online]. Available: \url{https://www.stat.berkeley.edu/~mmahoney/pubs/matrix3_SICOMP.pdf}
\BIBentrySTDinterwordspacing

\bibitem{achlioptas2007fast}
\BIBentryALTinterwordspacing
D.~Achlioptas and F.~McSherry, ``Fast computation of low-rank matrix approximations,'' \emph{Journal of the ACM}, vol.~54, no.~9, 2007. [Online]. Available: \url{https://www.cs.princeton.edu/courses/archive/spr04/cos598B/bib/Mcsherrysvd.pdf}
\BIBentrySTDinterwordspacing

\bibitem{mahoney2009cur}
\BIBentryALTinterwordspacing
M.~Mahoney and P.~Drineas, ``Cur matrix decompositions for improved data analysis,'' \emph{Proceedings of the National Academy of Sciences}, vol. 106, no.~3, 2009. [Online]. Available: \url{https://www.pnas.org/content/pnas/106/3/697.full.pdf}
\BIBentrySTDinterwordspacing

\bibitem{thurau2012deterministic}
\BIBentryALTinterwordspacing
C.~Thurau, K.~Kersting, and C.~Bauckhage, ``Deterministic cur for improved large-scale data analysis: An empirical study,'' in \emph{Proc. Int. Conf. on Data Mining}.\hskip 1em plus 0.5em minus 0.4em\relax SIAM, 2012. [Online]. Available: \url{https://epubs.siam.org/doi/pdf/10.1137/1.9781611972825.59}
\BIBentrySTDinterwordspacing

\bibitem{kannan2017randomized}
\BIBentryALTinterwordspacing
R.~Kannan and S.~Vempala, ``Randomized algorithms in numerical linear algebra,'' \emph{Acta Numerica}, vol.~26, 2017. [Online]. Available: \url{https://www.cc.gatech.edu/~vempala/papers/acta_survey.pdf}
\BIBentrySTDinterwordspacing

\bibitem{salakhutdinov2007probabilistic}
\BIBentryALTinterwordspacing
R.~Salakhutdinov and A.~Mnih, ``Probabilistic matrix factorization,'' in \emph{Proc. Advances in Neural Information Processing Systems (NeurIPS)}, 2007. [Online]. Available: \url{https://papers.nips.cc/paper/2007/file/d7322ed717dedf1eb4e6e52a37ea7bcd-Paper.pdf}
\BIBentrySTDinterwordspacing

\bibitem{zhang2011quantum}
G.~Zhang, ``Quantum-inspired evolutionary algorithms: A survey and empirical study,'' \emph{Journal of Heuristics}, vol.~17, 2011.

\bibitem{chia2020quantum}
N.~Chia \emph{et~al.}, ``Quantum-inspired algorithms for solving low-rank linear equation systems with logarithmic dependence on the dimension,'' in \emph{Int. Symp. on Algorithms and Computation (ISAAC)}, 2020.

\bibitem{gilyen2019quantum}
A.~Gilyen \emph{et~al.}, ``Quantum singular value transformation and beyond: Exponential improvements for quantum matrix arithmetics,'' in \emph{Proc. Symp. on the Theory of Computing (STOC)}.\hskip 1em plus 0.5em minus 0.4em\relax ACM, 2019.

\bibitem{chia2018quantum}
N.~Chia, H.-W. Lin, and C.-Y. Wang, ``Quantum-inspired sublinear classical algorithms for solving low-rank linear systems,'' \emph{arXiv:1811.04852 [cs.DS]}, 2018.

\bibitem{hornik1989multilayer}
K.~Hornik, M.~Stinchcombe, and H.~White, ``Multilayer feedforward networks are universal approximators,'' \emph{Neural networks}, vol.~2, no.~5, pp. 359--366, 1989.

\bibitem{rumelhart1986learning}
D.~E. Rumelhart, G.~E. Hinton, and R.~J. Williams, ``Learning representations by back-propagating errors,'' \emph{nature}, vol. 323, no. 6088, pp. 533--536, 1986.

\bibitem{lin2017does}
H.~W. Lin, M.~Tegmark, and D.~Rolnick, ``Why does deep and cheap learning work so well?'' \emph{Journal of Statistical Physics}, vol. 168, pp. 1223--1247, 2017.

\bibitem{coyleborn}
B.~Coyle, D.~Mills, V.~Danos, and E.~Kashefi, ``The born supremacy: quantum advantage and training of an ising born machine. npj quantum information, 6 (1), jul 2020.''

\bibitem{liu2018differentiable}
J.-G. Liu and L.~Wang, ``Differentiable learning of quantum circuit born machines,'' \emph{Physical Review A}, vol.~98, no.~6, p. 062324, 2018.

\bibitem{cheng2018information}
S.~Cheng, J.~Chen, and L.~Wang, ``Information perspective to probabilistic modeling: Boltzmann machines versus born machines,'' \emph{Entropy}, vol.~20, no.~8, p. 583, 2018.

\bibitem{allcock2018quantum}
\BIBentryALTinterwordspacing
J.~Allcock \emph{et~al.}, ``Quantum algorithms for feedforward neural networks,'' \emph{ACM Trans. on Quantum Computing}, 2018. [Online]. Available: \url{https://arxiv.org/abs/1812.03089v2}
\BIBentrySTDinterwordspacing

\bibitem{beer2020training}
\BIBentryALTinterwordspacing
K.~Beer \emph{et~al.}, ``Training deep quantum neural networks,'' \emph{Nature Communications}, vol.~11, 2020. [Online]. Available: \url{https://www.nature.com/articles/s41467-020-14454-2.pdf}
\BIBentrySTDinterwordspacing

\bibitem{mitarai2018quantum}
\BIBentryALTinterwordspacing
K.~Mitarai \emph{et~al.}, ``Quantum circuit learning,'' \emph{Physical Review A}, vol.~98, no.~3, 2018. [Online]. Available: \url{https://link.aps.org/doi/10.1103/PhysRevA.98.032309}
\BIBentrySTDinterwordspacing

\bibitem{verdon2019learning}
\BIBentryALTinterwordspacing
G.~Verdon \emph{et~al.}, ``Learning to learn with quantum neural networks via classical neural networks,'' \emph{arXiv:1907.05415 [quant-ph]}, 2019. [Online]. Available: \url{https://arxiv.org/abs/1907.05415v1}
\BIBentrySTDinterwordspacing

\bibitem{mcclean2018barren}
J.~R. McClean \emph{et~al.}, ``Barren plateaus in quantum neural network training landscapes,'' \emph{Nature Communications}, vol.~9, 2018.

\bibitem{zhu2019training}
D.~Zhu \emph{et~al.}, ``Training of quantum circuits on a hybrid quantum computer,'' \emph{Science Advances}, vol.~5, no.~10, 2019.

\bibitem{debnath2016demonstration}
S.~Debnath \emph{et~al.}, ``Demonstration of a small programmable quantum computer with atomic qubits,'' \emph{Nature}, vol. 536, 2016.

\bibitem{leyton2019robust}
V.~Leyton-Ortega, A.~Perdomo-Ortiz, and O.~Perdomo, ``Robust implementation of generative modeling with parametrized quantum circuits,'' \emph{Quantum Machine Intelligence}, vol.~3, 2020.

\bibitem{rossi2013quantum}
\BIBentryALTinterwordspacing
M.~Rossi \emph{et~al.}, ``Quantum hypergraph states,'' \emph{New J. of Physics}, vol.~15, no.~11, 2013. [Online]. Available: \url{https://doi.org/10.1088/1367-2630/15/11/113022}
\BIBentrySTDinterwordspacing

\bibitem{zhao2019quantum}
\BIBentryALTinterwordspacing
Z.~Zhao, J.~K. Fitzsimons, and J.~F. Fitzsimons, ``Quantum-assisted gaussian process regression,'' \emph{Phys. Rev. A}, vol.~99, no.~5, 2019. [Online]. Available: \url{https://link.aps.org/doi/10.1103/PhysRevA.99.052331}
\BIBentrySTDinterwordspacing

\bibitem{dayan1995helmholtz}
\BIBentryALTinterwordspacing
P.~Dayan \emph{et~al.}, ``The helmholtz machine,'' \emph{Neural Computation}, vol.~7, no.~5, 1995. [Online]. Available: \url{https://doi.org/10.1162/neco.1995.7.5.889}
\BIBentrySTDinterwordspacing

\bibitem{hinton1995wake}
\BIBentryALTinterwordspacing
G.~Hinton \emph{et~al.}, ``The “wake-sleep” algorithm for unsupervised neural networks,'' \emph{Science}, vol. 268, no. 5214, 1995. [Online]. Available: \url{https://www.science.org/doi/abs/10.1126/science.7761831}
\BIBentrySTDinterwordspacing

\bibitem{vandam2020hybrid}
T.~van Dam \emph{et~al.}, ``Hybrid helmholtz machines: A gate-based quantum circuit implementation,'' \emph{Quantum Information Processing}, vol.~19, 2020.

\bibitem{schuld2019quantum}
\BIBentryALTinterwordspacing
M.~Schuld and N.~Killoran, ``Quantum machine learning in feature hilbert spaces,'' \emph{Physical Review Letters}, vol. 122, no.~4, 2019. [Online]. Available: \url{https://link.aps.org/doi/10.1103/PhysRevLett.122.040504}
\BIBentrySTDinterwordspacing

\bibitem{zhao2019bayesian}
Z.~Zhao \emph{et~al.}, ``Bayesian deep learning on a quantum computer,'' \emph{Quantum Machine Intelligence}, vol.~1, 2019.

\bibitem{schuld2015simulating}
\BIBentryALTinterwordspacing
M.~Schuld, I.~Sinayskiy, and F.~Petruccione, ``Simulating a perceptron on a quantum computer,'' \emph{Physics Letters A}, vol. 379, no.~7, 2015. [Online]. Available: \url{https://www.sciencedirect.com/science/article/pii/S037596011401278X}
\BIBentrySTDinterwordspacing

\bibitem{bauckhage2017ising}
C.~Bauckhage \emph{et~al.}, ``Ising models for binary clustering via adiabatic quantum computing,'' in \emph{Proc. Int. Conf. on Energy Minimization Methods in Computer Vision and Pattern Recognition (EMMCVPR)}.\hskip 1em plus 0.5em minus 0.4em\relax Springer, 2017.

\bibitem{schoelkopf2002learning}
\BIBentryALTinterwordspacing
B.~Schölkopf and A.~Smola, \emph{Learning with Kernels: Support Vector Machines, Regularization, Optimization, and Beyond}.\hskip 1em plus 0.5em minus 0.4em\relax MIT Press, 2002. [Online]. Available: \url{https://ieeexplore.ieee.org/servlet/opac?bknumber=6267332}
\BIBentrySTDinterwordspacing

\bibitem{bauckhage2018adiabatic}
\BIBentryALTinterwordspacing
C.~Bauckhage \emph{et~al.}, ``Adiabatic quantum computing for kernel k=2 means clustering,'' in \emph{Proc. Conf. Learning, Knowledge, Data, Analytics (KDML-LWDA)}, 2018. [Online]. Available: \url{http://ceurws.org/Vol-2191/paper3.pdf}
\BIBentrySTDinterwordspacing

\bibitem{schuld2014quantum}
M.~Schuld, I.~Sinayskiy, and F.~Petruccione, ``Quantum computing for pattern classification,'' in \emph{Proc. Pacific Rim Int. Conf. on Artificial Intelligence (PRICAI)}.\hskip 1em plus 0.5em minus 0.4em\relax Springer, 2014.

\bibitem{Lloyd_2014_PCA}
\BIBentryALTinterwordspacing
S.~Lloyd, M.~Mohseni, and P.~Rebentrost, ``Quantum principal component analysis,'' \emph{Nature Physics}, vol.~10, no.~9, pp. 631--633, 7 2014. [Online]. Available: \url{https://doi.org/10.1038%2Fnphys3029}
\BIBentrySTDinterwordspacing

\bibitem{chen2020quantum}
D.~Chen, Y.~Xu, B.~Baheri, C.~Bi, Y.~Mao, Q.~Quan, and S.~Xu, ``Quantum-inspired classical algorithm for principal component regression,'' \emph{arXiv preprint arXiv:2010.08626}, 2020.

\bibitem{xie2017quantum}
Z.~Xie and I.~Sato, ``A quantum-inspired ensemble method and quantum-inspired forest regressors,'' in \emph{Asian Conference on Machine Learning}.\hskip 1em plus 0.5em minus 0.4em\relax PMLR, 2017, pp. 81--96.

\bibitem{zahorodko2021comparisons}
P.~V. Zahorodko \emph{et~al.}, ``Comparisons of performance between quantum-enhanced and classical machine learning algorithms on the ibm quantum experience,'' \emph{Journal of Physics: Conference Series}, vol. 1840, 2021.

\bibitem{havenstein2018comparisons}
C.~Havenstein, D.~Thomas, and S.~Chandrasekaran, ``Comparisons of performance between quantum and classical machine learning,'' \emph{SMU Data Science Review}, vol.~1, no.~4, 2018.

\bibitem{rebentrost2014quantum}
P.~Rebentrost, M.~Mohseni, and S.~Lloyd, ``Quantum support vector machine for big data classification,'' \emph{Physical Review Letters}, vol. 113, 2014.

\bibitem{wang2014application}
H.~K. Wang \emph{et~al.}, ``Application of the least squares support vector machine based on quantum particle swarm optimization for data fitting of small samples,'' \emph{Mechanical Science and Engineering IV}, vol. 472, 2014.

\bibitem{wittek2014supervised}
\BIBentryALTinterwordspacing
P.~Wittek, ``Supervised learning and support vector machines,'' in \emph{Quantum Machine Learning}.\hskip 1em plus 0.5em minus 0.4em\relax Academic Press, 2014. [Online]. Available: \url{https://www.sciencedirect.com/science/article/pii/B9780128009536000153}
\BIBentrySTDinterwordspacing

\bibitem{boser1992training}
\BIBentryALTinterwordspacing
B.~E. Boser, I.~Guyon, and V.~Vapnik, ``A training algorithm for optimal margin classifiers,'' in \emph{Proc. Conf. on Computational Learning Theory (COLT)}, 1992. [Online]. Available: \url{https://www.researchgate.net/profile/Bernhard-Boser/publication/2376111_A_Training_Algorithm_for_Optimal_Margin_Classifier/links/560eccc208ae0fc513ee8fc9/A-Training-Algorithm-for-Optimal-Margin-Classifier.pdf}
\BIBentrySTDinterwordspacing

\bibitem{QEM_Miyahara_2017}
\BIBentryALTinterwordspacing
H.~Miyahara, K.~Tsumura, and Y.~Sughiyama, ``Deterministic quantum annealing expectation-maximization algorithm,'' \emph{Journal of Statistical Mechanics: Theory and Experiment}, vol. 2017, no.~11, p. 113404, 11 2017. [Online]. Available: \url{https://dx.doi.org/10.1088/1742-5468/aa967e}
\BIBentrySTDinterwordspacing

\bibitem{QEM_kerenidis2020quantum}
I.~Kerenidis, A.~Luongo, and A.~Prakash, ``Quantum expectation-maximization for gaussian mixture models,'' 2020.

\bibitem{QEM_Miyahara_2020}
\BIBentryALTinterwordspacing
H.~Miyahara, K.~Aihara, and W.~Lechner, ``Quantum expectation-maximization algorithm,'' \emph{Physical Review A}, vol. 101, no.~1, 1 2020. [Online]. Available: \url{https://doi.org/10.1103%2Fphysreva.101.012326}
\BIBentrySTDinterwordspacing

\bibitem{QEM_7798981}
H.~Miyahara, K.~Tsumura, and Y.~Sughiyama, ``Relaxation of the em algorithm via quantum annealing for gaussian mixture models,'' in \emph{2016 IEEE 55th Conference on Decision and Control (CDC)}, 2016, pp. 4674--4679.

\bibitem{sergioli2018quantum}
G.~Sergioli, E.~Santucci, L.~Didaci, J.~A. Miszczak, and R.~Giuntini, ``A quantum-inspired version of the nearest mean classifier,'' \emph{Soft Computing}, vol.~22, no.~3, pp. 691--705, 2018.

\bibitem{poggiali2022quantum}
A.~Poggiali, A.~Berti, A.~Bernasconi, G.~Del~Corso, and R.~Guidotti, ``Quantum clustering with k-means: a hybrid approach,'' \emph{arXiv preprint arXiv:2212.06691}, 2022.

\bibitem{ning2023quantum}
T.~Ning, Y.~Yang, and Z.~Du, ``Quantum kernel logistic regression based newton method,'' \emph{Physica A: Statistical Mechanics and its Applications}, vol. 611, p. 128454, 2023.

\bibitem{kim2019quantum}
J.~S. Kim and C.~W. Ahn, ``Quantum algorithm on logistic regression problem,'' \emph{IEICE TRANSACTIONS on Information and Systems}, vol. 102, no.~4, pp. 856--858, 2019.

\bibitem{grant2018hierarchical}
E.~Grant, M.~Benedetti, S.~Cao, A.~Hallam, J.~Lockhart, V.~Stojevic, A.~G. Green, and S.~Severini, ``Hierarchical quantum classifiers,'' \emph{npj Quantum Information}, vol.~4, no.~1, p.~65, 2018.

\bibitem{liu2018design}
Z.~Liu, X.~Liang, and M.~Huang, ``Design of logistic regression health assessment model using novel quantum pso,'' in \emph{2018 IEEE 3rd International Conference on Cloud Computing and Internet of Things (CCIOT)}.\hskip 1em plus 0.5em minus 0.4em\relax IEEE, 2018, pp. 39--42.

\bibitem{khadiev2021quantum}
K.~Khadiev and L.~Safina, ``The quantum version of random forest model for binary classification problem,'' in \emph{CEUR Workshop Proc}, vol. 2842, 2021, pp. 30--35.

\bibitem{srikumar2022kernel}
M.~Srikumar, C.~D. Hill, and L.~C. Hollenberg, ``A kernel-based quantum random forest for improved classification,'' \emph{arXiv preprint arXiv:2210.02355}, 2022.

\end{thebibliography}

\end{document}